\documentclass[12pt]{article}
\usepackage{graphics}
\begin{document}

\title{Comment on "A new exactly solvable quantum model in $N$ dimensions" [Phys. Lett. A 375(2011)1431]}
\author{B. L. Moreno Ley and Shi-Hai Dong\thanks{Corresponding author. E-mail address:
dongsh2@yahoo.com;
Tel:+52-55-57296000 ext 55255; Fax: +52-55-57296000 ext 55015.}\\
{\footnotesize Departamento de F\'{\i}sica, Escuela Superior de
F\'{\i}sica y Matem\'{a}ticas, Instituto Polit\'ecnico Nacional, }\\
{\footnotesize Edificio 9, Unidad Profesional Adolfo L\'opez
Mateos, Mexico D. F. 07738, Mexico}}
\date{}
\maketitle

\begin{abstract}
We pinpoint that the work about "a new exactly solvable quantum
model in $N$ dimensions" by Ballesteros et al. [Phys. Lett. A {\bf
375} (2011) 1431] is not a new exactly solvable quantum model
since the flaw of the position-dependent mass Hamiltonian proposed
by them makes it less valuable in physics.

 \maketitle
\end{abstract}

In recent work \cite{angel-11}, the authors Ballesteros et al.
claimed that they have found a new exactly solvable quantum model
in $N$ dimensions given by
\begin{equation}\label{}
H=-\frac{\hbar^2}{2(1+\lambda r^2)}\nabla^2+\frac{\omega^2
r^2}{2(1+\lambda r^2)},
\end{equation} where we prefer to use variable $r$ instead of original one
$q$ for convenience.

They found that the spectrum of this model is shown to be
hydrogen-like (should be harmonic oscillator-like), and their
eigenvalues and eigenfunctions are explicitly obtained by
deforming appropriately the symmetry properties of the
$N$-dimensional harmonic oscillator. It should be pointed out that
such treatment approach is incorrect since the kinetic energy term
should be defined as \cite{chen04}
\begin{equation}\label{12-nabla-2}
\nabla_N\frac{1}{m(r)}\nabla_N\psi({\bf
r})=\left(\nabla_N\frac{1}{m(r)}\right)\cdot[\nabla_N \psi({\bf
r})]+\frac{1}{m(r)}\nabla_N^2\psi({\bf r}).
\end{equation}
For $N$-dimensional spherical symmetry, we take the
wavefunctions\index{Wavefunction} $\psi({\bf r})$ as follows
\cite{dong02}:
\begin{equation}\label{mass-radial}
\psi({\bf r})=r^{-(N-1)/2}R(r)Y_{l_{N-2, \ldots, l_1}}^{l}({\bf
\hat x}).
\end{equation}
Substituting this into the position-dependent effective mass
Schr\"{o}dinger equation
\begin{equation}\label{mass-sch}
\nabla_N\left(\frac{1}{m(r)}\nabla_N\psi({\bf
r})\right)+2[E-V(r)]\psi({\bf r})=0,
\end{equation}
allows us to obtain the following radial position-dependent mass
Schr\"{o}dinger equation in arbitrary dimensions
\begin{equation}\label{mass-radial-1}
\left\{\frac{d^2}{dr^2}+\frac{m'(r)}{m(r)}\left(\frac{N-1}{2r}-\frac{d}{dr}\right)-\frac{\eta^2-1/4}{r^2}
+2m(r)[E-V(r)]\right\}R(r)=0,
\end{equation}
where $m(r)=(1+\lambda r^2), m'(r)=dm(r)/dr$ and $\eta=|l-1+N/2|$.
Since the operator $\nabla_{N}$ does not commutate with the
position-dependent mass $m(r)$, then this system does not exist
exact solutions at all. This can also be proved unsolvable to
Eq.(\ref{mass-radial-1}) if substituting the position-dependent
mass $m(r)$ into it.

On the other hand, the choice of the position-dependent mass
$m(r)$ has no physical meaning since the mass $m(r)$ goes to
infinity when $r\rightarrow\infty$. Moreover, it is shown from
Eq.(1) that the position-dependent mass $m(r)$ in kinetic term is
equal to $(1+\lambda r^2)$, but it was taken as $1/(1+\lambda
r^2)$ for the harmonic oscillator term. Accordingly, the wrong
expression of the Hamiltonian in position-dependent mass
Schr\"{o}dinger equation in arbitrary dimensions $N$, the flaw of
the chosen position-dependent mass $m(r)$ as well as its
inconsistence between the kinetic term and the harmonic oscillator
term make it less valuable in physics.

\vskip 5mm {\bf \Large Acknowledgments}: This work was supported
partially by 20110491-SIP-IPN, COFAA-IPN, Mexico.

\end{document}